\documentclass[english]{IEEEtran}
\usepackage[T1]{fontenc}
\usepackage[latin9]{inputenc}
\usepackage{amsmath}
\usepackage{amssymb}
\usepackage{esint}

\makeatletter

\providecommand{\tabularnewline}{\\}

\usepackage{url}

\makeatother

\begin{document}

\title{A mixed precision semi-Lagrangian algorithm and its performance on accelerators}
\author{Lukas Einkemmer%
\thanks{L. Einkemmer is with the University of Innsbruck, Austria}%
\thanks{E-mail: lukas.einkemmer@uibk.ac.at}%
%\thanks{This work is  supported by the Austrian Science Fund (FWF) -- project id: P25346.}%
}

\markboth{The 2016 International Conference on High Performance Computing \& Simulation}{A mixed precision semi-Lagrangian algorithm}

\maketitle 

\begin{abstract}
In this paper we propose a mixed precision algorithm in the context of the semi-Lagrangian discontinuous Galerkin method. The performance of this approach is evaluated on a traditional dual socket workstation as well as on a Xeon Phi and a NVIDIA K80. We find that the mixed precision algorithm can be implemented efficiently on these architectures. This implies that, in addition to the considerable reduction in memory, a substantial increase in performance can be observed as well. Moreover, we discuss the relative performance of our implementations.  

\end{abstract}

\IEEEpeerreviewmaketitle

\section{Introduction}

Semi-Lagrangian methods are an important class of numerical algorithms
to solve hyperbolic partial differential equations. These methods
do not suffer from a Courant--Friedrichs--Lewy (CFL) condition and
can be applied without solving a linear system of equations. Consequently,
semi-Lagrangian methods are computationally attractive and have been
applied to a wide variety of problems ranging from weather forecasting
to plasma simulations. 

Due to the prevalence of these algorithms in applications, studying
semi-Lagrangian methods and their efficient implementation is an important
research area. In particular, in plasma physics applications problems
in a high dimensional setting (up to 6 dimensions; three position
and three velocity directions) are common. Thus, both memory consumption
and computational performance are a major concern in such simulations.

Let us note that contrary to the much studied stencil codes, these
algorithms, in general, do not have pre-determined memory access patterns
and in some cases have non-uniform degrees of freedom (i.e.~the specific
computation performed depends on index of the data under consideration).
This potentially presents a problem for accelerators as aligned data
access is usually required in order to obtain optimal performance.

The crucial part in any semi-Lagrangian algorithm is the interpolation
or projection step. In this step the translated function is interpolated/projected
back to the grid/approximation space. In the literature a number of
numerical schemes have been introduced. Most commonly, cubic spline
interpolation is employed in practice. However, recently the so-called
semi-Lagrangian discontinuous Galerkin scheme has been considered.
This method is competitive with spline interpolation but is a completely
local numerical method. This fact greatly facilitates the implementation
on parallel architectures (both on shared as well as on distributed
memory systems). 

Due to the constraints outlined above and since the algorithm is memory
bound, implementing it using single precision floating point arithmetic
would significantly improve performance as well as reduce memory consumption.
This is particularly true on accelerators, such as graphic processing
units (GPUs) and the Intel Xeon Phi, for which memory is a more scarce
resource compared to traditional central processing unit (CPU) based
systems. In addition, since the memory available per core will most
likely continue to decrease on future architectures, the problem of
memory scarcity is going to become an even bigger issue in the future.

However, since in many practical simulations a large number of time
steps, and consequently a large number of projections, have to be
performed, conservation of mass at least up to double precision accuracy
has been the gold standard in the physics community. This is particularly
important as for the classic time-splitting approach conservation
of mass implies conservation of momentum as well as a range of other
desirable properties.

Mixed precision algorithms have attracted some interest recently.
This is particularly true for numerical linear algebra algorithms
(see, for example, \cite{baboulin2009,emans2010,lei2010,glimberg2013,goddeke2011,buttari2007}).
For example, in the context of iterative methods it is very natural
to compute a first approximation in single precision which is then
refined, if necessary, using double precision arithmetics. Also computing
the preconditioner in single precision to increase performance is
an often used technique. Furthermore, mixed precision algorithms for
some specific applications have been proposed as well (see, for example,
\cite{glimberg2013,le2013}) and even the possibility to automatically
convert certain parts of computer programs to single precision has
been investigated (see \cite{lam2013}). However, the traditional
formulation of semi-Lagrangian methods poses significant difficulties
if a mixed precision implementation is to be considered.

In this paper we propose a mixed precision semi-Lagrangian approach
that succeeds in conserving the mass up to double precision while
storing almost all of the data used in the computation as single precision
floating point numbers. We use the before mentioned semi-Lagrangian
discontinuous Galerkin method, but instead of storing certain function
evaluations at a non-equidistant grid (as has been the predominant
approach in the literature), we use the fact that the corresponding
approximation can be written as an expansion in the Legendre polynomials.
The first coefficient in this expansion is simply the mass in a subset
of the computational domain (which is then stored in double precision).
Furthermore, we will investigate the performance of this algorithm
on an Intel Xeon Phi and a NVIDIA K80 GPU and compare the performance
attained to a dual socket workstation.

The remainder of this paper is structured as follows. In section \ref{sec:background}
we discuss the problem as well as the algorithm proposed to solve
it in more detail. In addition, we briefly discuss the hardware architectures
used in our implementation. In section \ref{sec:results} we study
the error of the proposed algorithm, the different implementations,
and our findings with respect to performance. These can be summarized
as follows.
\begin{itemize}
\item The semi-Lagrangian discontinuous Galerkin scheme can be implemented
efficiently using mixed precision arithmetics, resulting in both a
significant memory reduction as well as a significant performance
increase. This is true across all the hardware architectures considered
here (dual socket workstation, Intel Xeon Phi, and NVIDIA K80 GPU).
\item For traditional CPU based systems our implementation, in most of the
cases considered, is close to the theoretical attainable performance
for memory bound problems on that architecture.
\item For the Xeon Phi we observe a speed up of approximately 50\% compared
to the CPU implementation (using the code that has been optimized
for the CPU system). Thus, in this case the Xeon Phi can act as a
drop-in replacement. 
\item For the K80 we observe a speedup of approximately $2.5$ compared
to the CPU implementation. This corresponds to 75\% of the peak performance
for that architecture. 
\end{itemize}
Finally, we conclude in section \ref{sec:conclusion}.

\section{Background \& Motivation\label{sec:background}}

The Vlasov equation (here stated in 1+1 dimensions)
\[
\partial_{t}f(t,x,v)+v\partial_{x}f(t,x,v)+E(x)\partial_{v}f(t,x,v)=0
\]
models the time evolution of a plasma system. The sought after quantity
is the particle density function $f(t,x,v)$, and the electric field
is denoted by $E(x)$. This model is important for the description
of non thermalized plasmas; i.e.~ plasmas where fluid models (such
as magnetohydrodynamics) are not applicable. In the seminal paper
by Cheng \& Knorr \cite{cheng1976} it was recognized that by applying
a time splitting approach, the Vlasov equation is reduced to a sequence
of one-dimensional advections. That is, in order to solve the Vlasov
equation efficiently it is necessary to develop a good integrator
for
\begin{equation}
\partial_{t}u(t,x)+a\partial_{x}u(t,x)=0,\label{eq:advection}
\end{equation}
where $a\in\mathbb{R}$ is a constant. Let us emphasize that the simplicity
of this equation (for which even an analytic solution can be derived)
is deceiving. In fact, a large body of literature has been devoted
to this problem (see, for example, \cite{sonnendrucker1999,filbet2003,crouseilles2011,rossmanith2011,qiu2011}).
For solving the Vlasov equation it is important that the numerical
scheme is free of a CFL condition. Otherwise, the numerical scheme
would be forced to take excessively small time steps. Therefore, so-called
semi-Lagrangian methods have become popular in this field. These methods
follow the characteristics backward in time in order to compute the
function values at the grid points. Note, however, that the feet of
the characteristics do not necessarily coincide with the grid. Thus,
an interpolation procedure has to be employed. Note that in the case
where the characteristics can be determined analytically (as is the
case for equation (\ref{eq:advection})) this is the only approximation
made (except for the time splitting). 

Consequently it is important to choose a suitable interpolation scheme.
An obvious choice is to use local polynomial interpolation or methods
based on fast Fourier techniques. Especially the latter was used quite
extensively in many Vlasov simulations (see, for example, \cite{cheng1976,klimas1994}).
However, more recently interpolation using cubic splines has become
the de facto standard \cite{sonnendruecker2011} and the performance
of numerical software packages that implement such methods has been
investigated in some detail (see, for example, \cite{rozar2013,bigot2013,grandgirard}).

This is due to the fact that spline interpolation is mass conservative
and shows little numerical diffusion compared to alternative approaches.
In addition, it is not as prone to oscillations (and thus to the appearance
of negative values) as are Fourier techniques.

However, the procedure also suffers from a number of shortcomings.
Most notably that it is a global algorithm. That is, the construction
of the cubic spline (for which we have to solve a sparse linear system)
couples each degree of freedom with each other degree of freedom.
The resulting all-to-all communication is a serious issue with respect
to the scalability of such algorithms.

\subsection{Description of the algorithm}

In recent years an alternative method has emerged. The so-called semi-Lagrangian
discontinuous Galerkin scheme was independently proposed by \cite{crouseilles2011,rossmanith2011,qiu2011}.
Since a variant of this method will be used in the present paper,
we will describe it in some detail. First, the computational domain
is divided into a number of cells $C_{i}=[x_{i-1/2},x_{i+1/2}]$.
In each of these cells we approximate the exact solution by a polynomial
of degree $p$. This approximation requires the storage of $o=p+1$
degrees of freedom. Let us note that no continuity constraint is enforced
at the cell boundaries. Thus, the approximation of the continuous
solution $u(t,x)$ is performed by a function with discontinuities
at the cell interfaces. The corresponding jumps are bounded in magnitude
by the discretization error. To solve equation (\ref{eq:advection})
we translate the approximant (this can be done analytically) and then
perform an $L^{2}$ projection to the subspace of piecewise polynomials
up to degree $p$. This can be easily accomplished by choosing an
orthogonal basis of the polynomial approximation space and results
in an approximation of order $o$.

The semi-Lagrangian discontinuous Galerkin method is mass conservative
by construction and compares very favorably to the cubic spline interpolation
(see, for example, \cite{crouseilles2011,einkemmer2016,mehrenberger2013,einkemmer2015error}).
In addition, it requires at most the data from two adjacent cells
in order to compute the approximation at the subsequent time step
(note that this behavior is independent of the CFL number). Thus,
the semi-Lagrangian discontinuous Galerkin method is a completely
local scheme. This greatly facilitates the implementation on distributed
memory systems (see, for example, \cite{einkemmer2015}).

In most of the literature, the Lagrange basis polynomials at the Gauss--Legendre
quadrature nodes are used as the basis. This has the advantage that
the degrees of freedom are function evaluations on a non-equidistant
grid. However, since the degrees of freedom correspond to function
evaluations, even for smooth functions, all of them are approximately
equal in magnitude. Thus, this representation is unsuitable in the
context of a mixed precision implementation.

In order to remedy this issue we use the fact that the first $o$
Legendre polynomials (appropriately scaled to the corresponding cell
size) form a basis of the space of polynomials up to degree $p$.
Thus, in order to approximate $u(x)$ in the cell $[-h/2,h/2]$ we
use
\[
u(x)\approx\sum_{j=0}^{p}c_{j}P_{j}(\tfrac{2}{h}x),
\]
where $P_{j}$ is the $j$th Legendre polynomial defined on the interval
$[-1,1]$. The degrees of freedom, which are consequently stored in
computer memory, are the coefficients $c_{j}$ that appear in the
expansion. These coefficients can be computed as follows
\begin{equation}
c_{j}=\frac{2j+1}{2}\frac{2}{h}\int_{-h/2}^{h/2}u(x)P_{j}(\tfrac{2}{h}x)\,\mathrm{d}x.\label{eq:cj}
\end{equation}
This integral is easily solved exactly by performing a Gauss--Legendre
quadrature. The advantage of this representation is that for a smooth
function $u$ we have $c_{j}\sim h^{j}$. Thus, $c_{0}$ will be $\mathcal{O}(1)$
while higher order coefficients become progressively smaller. This
scaling can be exploited by storing the higher order coefficients
with less precision (i.e.~ using single precision). This can be interpreted
as a, in general lossy, compression scheme where less important coefficients
are stored using fewer bytes.

Let us also note that $c_{0}$ corresponds to the mass in $[-h/2,h/2]$.
This is especially convenient for a mixed precision implementation
as storing $c_{0}$ in double precision (and the remaining coefficients
in single precision) automatically ensures conservation of mass up
to double precision accuracy. 

In the present text we will not derive the implemented algorithm in
any detail. What is important for the present discussion, however,
is that by performing the Gauss--Legendre quadrature of equation (\ref{eq:cj})
we obtain the following update rule for the coefficient $c_{ij}^{n+1}$
at time $t_{n+1}$ ($i$ is the cell index) 
\[
c_{ij}^{n+1}=\sum_{l}A_{jl}c_{i^{\star}l}^{n}+\sum_{j}B_{jl}c_{i^{\star}+1;l}^{n}
\]
where $A\in\mathbb{R}^{o\times o}$ and $B\in\mathbb{R}^{o\times o}$
are matrices that only depend on the time step size and $i^{\star}$
is the integer part of the CFL number. Note that the CFL number is
not a fixed quantity but can depend, for example, on the electric
field. Thus, it is not a-priori known by how much the index $i$ is
translated in order to obtain $i^{\star}$. 

The matrices can be precomputed at the beginning of each time step
and thus their cost will not be a major concern except for an extremely
coarse space discretization. Furthermore, the size of the matrices
is negligible in comparison to the cache size and it is thus reasonable
to assume that access to these matrices is quick. Consequently, the
algorithm requires (assuming a perfect cache) one memory load and
one memory store per degree of freedom, which we have to compare to
the $4o-1$ arithmetic instructions. Thus, even for a sixth-order
method the flop/byte ratio is only about $1.4$ for double precision
and $2.9$ for single precision. This is significantly below the flop/byte
ratio for all the computer systems considered here (see Table \ref{tab:hw-config}).

\begin{table}[h]
\caption{Hardware characteristics of the dual socket workstation used in the
numerical simulations. Peak arithmetic performance for single and
double precision and the attainable memory bandwidth as well as the
flop/byte ratios are listed. The dagger indicates that the bandwidth
achieved by a memory copy benchmark is listed (and not the theoretical
bandwidth specified by the vendor). \label{tab:hw-config}}

\centering{}%
\begin{tabular}{rrrrrrrr}
\hline 
 &  & \multicolumn{2}{c}{TFlops/s} &  &  & \multicolumn{2}{c}{flop/byte}\tabularnewline
\cline{3-4} \cline{7-8} 
 &  & double & single &  & GB/s & double & single\tabularnewline
\hline 
2x E5-2630 v3 &  & 0.6 & 1.2 &  & 59 & 10.2 & 20.3\tabularnewline
1x Xeon Phi 7120 &  & 1.2 & 2.4 &  & $\text{150}^{\dagger}$ & 8 & 16\tabularnewline
0.5x K80 &  & 1.5 & 4.4 &  & $\text{170}^{\dagger}$ & 8.8 & 25.9\tabularnewline
\hline 
\end{tabular}
\end{table}

However, floating point arithmetic is not the only concern. Since
the algorithm requests a memory load for the addresses $i^{\star}$
and $i^{\star}+1$ in one iteration and to $i^{\star}+1$ and $i^{\star}+2$
in the next, cache performance is vital to obtain good performance.
This is especially true for the CPU and Intel Xeon Phi implementation
as in those instances there is little possibility to directly manipulate
the cache. We will discuss a number these issues in more detail in
the next section.

\subsection{Hardware architecture}

As has been mentioned before, reducing memory consumption is especially
important on accelerators due to the limited amount of memory available
on these systems. In the present paper we will use the Intel Xeon
Phi 7120 which includes 16 GB of memory and the NVIDIA K80 which consists
of two GPUs in a single package with 12 GB of memory each. The purpose
of the present section is to describe these two architectures.

Let us start with the Intel Xeon Phi due its similarity with the standard
x86 architecture. In fact, on the Intel Xeon Phi a full Linux operating
system is employed. The Xeon Phi can either be used in offload mode
(where the CPU and the Xeon Phi work together and exchange data over
the PCIe bus) and in standalone mode (where a program runs on the
Xeon Phi without intervention from the CPU system). The preferred
programming model for the Xeon Phi is OpenMP which we use in all our
implementations. However, they are also important differences. The
Intel Xeon Phi (similar to GPUs) uses GDDR5 memory which is optimized
for applications which can exploit high memory bandwidth. However,
it should be emphasized that to obtain this bandwidth certain access
pattern are required. This has to be contrasted with DDR3 memory which
is also fairly effective if random access patterns are used. The Xeon
Phi consists of 61 cores each of which has access to an 32 KB L1 cache
and an 512 KB L2 cache. The latter can interchange data via a ring
interconnect. Let us also mention that the Xeon Phi implements eight
double wide vector units (AVX512).

On the other hand, the NVIDIA K80 is a more lightweight architecture.
No operating system runs on the card and thus computations on the
GPU always have to be controlled by an application running on the
host system. Similar to the Xeon Phi, the K80 uses GDDR5 memory. Although
caches are a more recent addition to GPUs, this Kepler generation
card includes both an L2 cache and an L1 cache. The latter is implemented
together with a so-called shared memory. Shared memory is basically
an L1 cache that can be directly controlled by the program (thus it
is a user managed cache). The relative size of the L1 cache and the
shared memory can be configured to some extend and these two types
of memory are shared by all blocks running on the same streaming multiprocessor.
The K80 exposes significantly more parallelism with 2496 so-called
CUDA cores. Note, however, that 32 such CUDA cores are collected in
one warp and all threads in a particular warp have to execute the
same instruction if optimal performance is to be achieved. Thus, one
might argue that the GPU consists of 78 cores with 32 threads grouped
together in a vector unit. The difference to the vector units found
more commonly in CPUs, however, is the programming model. While for
CPUs/Xeon Phi vectorization is mostly delegated to the compiler, the
CUDA programming model exposes this behavior more directly to the
programmer. An additional difference is that in principle any command
can be vectorized, as long as all threads within a warp execute the
same statement (i.e.~there is no branch divergence). Since CUDA is
the most common programming model for NVIDIA GPUs, we use it in all
our implementations.

\section{Results\label{sec:results}}

\subsection{Accuracy}

It is well accepted in the Vlasov community that preserving mass up
to machine precision (i.e.~ up to double precision accuracy) is vital
in order to obtain a physically reasonable solution (especially for
long time integration). This is perhaps the most serious argument
against using single precision floating point numbers in such simulations.
However, as mentioned before, to only store the coefficient $c_{0}$
in double precision (and the remaining coefficients in single precision)
is sufficient in order to retain this behavior.

The purpose of the present section is to confirm this result numerically
and to investigate the different numerical schemes that result as
a consequence of increasing the number of coefficients stored in double
precision. To that end a number of numerical results are listed in
Table \ref{tab:accuracy}. We can easily observe the expected behavior
that all numerical methods which store at least $c_{0}$ in double
precision conserve the mass up to that accuracy.

\begin{table}
\caption{The error (compared to the double precision implementation) and the
error in mass is shown for a smooth initial value (top) and an oscillatory
initial value (bottom). In both cases $10^{4}$ time steps are conducted
and the error is measured in the discrete $L^{2}$ norm. The number
of coefficients that are stored in double precision is given in the
second column. \label{tab:accuracy}}

\centering
\begin{tabular}{ccrr}
order & \# double & error & error (mass) \\
\hline
  2  & 1 &  8.98e-10 &   4.44e-15\\
     & 0 &  1.09e-05 &   2.56e-06\\
&& \\
4  & 3  &  6.16e-14 &   1.21e-14\\
   & 2  &  6.41e-13 &   1.42e-14\\
   & 1  &  3.55e-10 &   6.15e-15\\
   & 0  &   2.80e-05 &   1.31e-05
\end{tabular}

$\phantom{A}$

\begin{tabular}{ccrr}
order & \# double & error & error (mass) \\
\hline
2 & 1 &   9.22e-10 &  1.04e-14\\
  & 0 &   5.98e-06 &  2.36e-06\\
  &   &            & \\
4 & 3 &   1.22e-12 &  1.11e-14\\
  & 2 &   6.32e-10 &  1.06e-14\\
  & 1 &   6.05e-08 &  1.64e-14\\
  & 0 &   6.54e-05 &  9.81e-06\\
\end{tabular} 
\end{table}

Let us now discuss the error in the numerical solution that is introduced
by storing some coefficients in single precision. This error is computed
by comparing the numerical scheme under investigation to the double
precision implementation. The corresponding results are given in Table
\ref{tab:accuracy}. We observe that for the smooth initial value
storing $c_{0}$ in double precision (and all other coefficients in
single precision) already yields very good accuracy. 

For the oscillatory initial value storing more coefficients in double
precision has some effect. Note, however, that storing only $c_{0}$
in single precision already yields a three order of magnitude improvement
as well as ensures mass conservation up to double precision accuracy.
Let us also remark that the error committed by the numerical approximation
(i.e.~by the projections performed in course of the time integration)
are large compared to the roundoff errors (except for the single precision
implementation).

From a practical point of view it seems that there are few reasons
to store more than one or two coefficients in double precision. Thus,
in the following sections we will mostly focus on the numerical scheme
were either $c_{0}$ is stored in double precision (for the second
and fourth order scheme) and the scheme where $c_{0}$ and $c_{1}$
are stored in double precision (for the fourth order scheme).

To conclude this section let us note that storing only $c_{0}$ in
memory results in a reduction in memory by a factor of $1.6$. Note,
however, that this effect becomes more pronounced in higher dimensions.
For example, in two dimensions the reduction in memory is already
a factor of $1.8$ and in dimension three a factor of $1.97$. The
latter is certainly almost indistinguishable from a pure single precision
implementation, where we obtain a reduction by a factor of $2$. This
should be kept in mind in the following discussion, even though for
simplicity we only consider the one-dimensional case in this paper.

\subsection{CPU performance}

As described in section \ref{sec:background} the algorithm under
consideration is memory bound. Thus, we use the achieved bandwidth
(measured in GB/s) as the metric of performance. 

The numerical results for the CPU implementation are shown in Table
\ref{tab:CPU-performance}. If we compare the bandwidth achieved to
the theoretical bandwidth of that architecture (59 GB/s), we find
that the algorithm performs very close to that limit. Thus, the implementation
is almost optimal for both the second and the fourth order method.
Let us note, however, that especially for the fourth order method
we observe a slight decrease in performance as we increase the number
of coefficients that are stored in single precision.

\begin{table}
\caption{The performance attained by the CPU implementation is shown for the
second and the fourth order scheme. The number of coefficients that
are stored in double precision is listed for each configuration. The
table gives the achieved speedup and the reduction in memory compared
to the double precision implementation. \label{tab:CPU-performance}}

\centering
\begin{tabular}{ccrrr}
order & \# double & bandwidth & speedup & memorydown \\
\hline
2 & 2 & 55.2 GB/s & --   & --  \\
2 & 1 & 51.3 GB/s & 1.24 & 1.33 \\
2 & 0 & 51.9 GB/s & 1.88 & 2.00 \\
&&&& \\
4 & 4 & 52.8 GB/s & -- & --\\
4 & 2 & 48.5 GB/s & 1.22 & 1.33\\
4 & 1 & 42.2 GB/s & 1.28 & 1.60\\
4 & 0 & 43.4 GB/s & 1.64 & 2.00\\
\end{tabular}
\end{table}

Let us note that in the present implementation the order of the method
and the number of coefficients that are stored in single precision
is available at compile time. This gives the compiler more information
for optimization. If this is not done (i.e.~if a generic implementation
is considered) the performance is cut approximately in half.

\subsection{Xeon Phi performance}

The Xeon Phi implementation is identical to the implementation that
was run on the dual socket workstation in the previous section. For
the Xeon Phi it is even more important to avoid a generic implementation
in favor of making all the relevant constants available at compile
time. All the simulations conducted here are run in standalone mode.

The performance results are shown in Figure \ref{tab:Phi-performance}.
The Xeon Phi manages to outperform the CPU based system by approximately
30 to 80\% (depending on the configuration). This variance in speedup
is mostly due to the performance characteristics of the CPU system
(the performance across the different configurations is actually more
consistent on the Xeon Phi). For the important configuration $o_{1}=1$,
$o_{2}=3$ it is observed that the corresponding implementation outperform
the double precision implementation by 80\% with a reduction in memory
by 60\%. 

\begin{table}
\caption{The performance attained by the Xeon Phi implementation is shown for
the second and the fourth order scheme. The number of coefficients
that are stored in double precision is listed for each configuration.
The table gives the achieved speedup and the reduction in memory compared
to the double precision implementation. \label{tab:Phi-performance}}

\centering
\begin{tabular}{ccrrr}
order & \# double & bandwidth & speedup & memorydown \\
\hline
2 & 2 & 76.4 GB/s & --   & --  \\
2 & 1 & 67.3 GB/s & 1.17 &  1.33 \\
2 & 0 & 76.2 GB/s & 1.99 &  2.00 \\
&&&& \\
4 & 4 & 67.9 GB/s & -- & --\\
4 & 2 & 72.0 GB/s & 1.41 & 1.33\\
4 & 1 & 76.3 GB/s & 1.80 & 1.60\\
4 & 0 & 68.6 GB/s & 2.02 & 2.00\\
\end{tabular}
\end{table}

Let us note that we achieve only slightly more than one-half of the
theoretical performance possible on the Xeon Phi. To improve these
results would most likely require a more sophisticated implementation
specifically tuned to the architecture of the Xeon Phi. It can be
argued, however, that obtaining a substantial performance improvement
with the same code that is used on the CPU is one of the major selling
points of the Xeon Phi. Although this is only occasionally the case,
it holds true for the implementation described here.

\subsection{GPU performance}

The GPU code described in this section has been implemented using
the CUDA framework. In this programming paradigm most of the specifics
of the GPU architecture are exposed to the programmer (although there
are some major exceptions). In particular, it is the programmers responsibility
to specify how much data each thread is given to process. As a starting
point we have ported our CPU implementation to the GPU. That is, each
thread is responsible for a single degree of freedom. This obviously
requires some branching statements in case of the mixed precision
implementation. However, since the performance of the algorithm is
dictated by memory bandwidth, loosing a few cycles due to warp divergence
is not a major concern. The achieved performance of the resulting
code is slightly above 60 GB/s. In some sense this is disappointing
as only one-third of the bandwidth available on the GPU is exploited.
However, we should emphasize that this is still a fully generic code;
i.e.~ the order of the method is not known at compile time and only
one kernel is used. If we do the same on the CPU we achieve approximately
35 GB/s. Thus, we still observe some speedup on the GPU if naive implementations
on both platforms are compared. In fact, the GPU implementation is
even slightly faster than what is theoretically possible on the CPU.
The situation is even more dramatic on the Xeon Phi, where the generic
implementation achieves only a performance below 15 GB/s.

An investigation has shown that the performance of the described GPU
implementation is mostly limited by cache misses. Fortunately, the
CUDA architectures gives us direct access to shared memory which we
use to cache all memory loads. The performance of the resulting implementation
is shown in Figure \ref{tab:GPU-performance}. 

\begin{table}
\caption{The performance attained by the shared memory GPU implementation is
shown for the second and the fourth order scheme. The number of coefficients
that are stored in double precision is listed for each configuration.
The table gives the achieved speedup and the reduction in memory compared
to the double precision implementation. \label{tab:GPU-performance}}

\centering
\begin{tabular}{ccrrr}
order & \# double & bandwidth & speedup & memorydown \\
\hline
2 & 2 & 110.1 GB/s & --   & --  \\
2 & 1 &  89.9 GB/s & 1.09 &  1.33 \\
2 & 0 &  90.9 GB/s & 1.65 &  2.00 \\
&&&& \\
4 & 4 & 95.0 GB/s & -- & --\\
4 & 2 & 72.1 GB/s & 1.01 & 1.33\\
4 & 1 & 60.0 GB/s & 1.01 & 1.60\\
4 & 0 & 73.2 GB/s & 1.54 & 2.00\\
\end{tabular}
\end{table}

We observe 95.6 GB/s for the double precision implementation (for
the fourth order scheme) and 109.7 GB/s for the second order scheme.
This is a distinctive improvement. However, the performance of the
mixed precision and the single precision implementation are significantly
worse in terms of the achieved memory bandwidth. What we have to conclude
is that while the mixed precision implementation saves memory, it
does not improve the performance of the computation.

The reason for the disappointing performance of the mixed precision
implementation is that since each thread only processes a single degree
of freedom, the overhead for the necessary index calculations is relatively
high. This is particularly a problem for the present algorithm as,
contrary to say a finite difference stencil, it is important for the
subsequent calculation at what location the currently processed degree
of freedom resides within a given cell. Thus, the index calculations
are more involved than is usually the case and can consequently not
be completely hidden. This effect is more pronounced for the mixed
precision and single precision implementation as less time is required
to read and write to memory (which aggravates the problem of hiding
this overhead).

The natural question then is how to improve the performance of this
implementation. We have accomplished this by making each thread responsible
for a single cell (not a single degree of freedom). This makes the
index calculation easier and alleviates the overhead as more work
is done per thread. We should note, however, that the resulting code
is specific to each configuration. That is, one separate kernel has
to be written for each configuration of the method (similar to the
CPU and Xeon Phi implementation). As we can see from the results in
Table \ref{tab:GPU2-performance}, the performance is dramatically
improved across the board. Moreover, independent of the implementation
we observe approximately 130 GB/s where the penalty for the mixed
precision implementation is below 10\%. Thus, the mixed precision
implementation not only decreases the memory consumption but also
increases the performance by a similar factor.

\begin{table}
\caption{The performance attained by the improved GPU implementation is shown
for the fourth order scheme. The number of coefficients that are stored
in double precision is listed for each configuration. The table gives
the achieved speedup and the reduction in memory compared to the double
precision implementation.\label{tab:GPU2-performance}}

\centering
\begin{tabular}{ccrrr}
order & \# double & bandwidth & speedup & memorydown \\
\hline
4 & 4 & 137.9 GB/s & -- & --\\
4 & 1 & 130.1 GB/s & 1.51 & 1.60\\
4 & 0 & 142.5 GB/s & 2.07 & 2.00\\
\end{tabular}
\end{table}

To conclude this section, let us discuss what limits the performance
of the present implementation. First, non aligned memory access is,
to some extend, unavoidable in the present numerical scheme (as the
data inside each cell is not a multiple of $128$ byte). This results
in a reduction of performance by approximately 10\%. Second, storing
data in shared memory requires a block-wide synchronization barrier.
This adds an additional overhead of approximately 10\%. Finally, the
remaining 10\% of performance loss are attributed to the fact that
the required computations can not be perfectly interleaved with the
memory accesses. In total our kernels achieve approximately 75\% of
the peak performance on the K80.

\section{Conclusion\label{sec:conclusion}}

We have demonstrated that mixed precision semi-Lagrangian algorithms
can be implemented in an efficient manner. On all the computer architectures
considered (CPU, Xeon Phi, GPU) significant speedups can be observed
in addition to the reduction in memory. Thus, we conclude that a mixed
precision of the semi-Lagrangian discontinuous Galerkin method is
a viable numerical scheme for simulations on both traditional CPU
based systems as well as on modern accelerators.

\bibliographystyle{plain}
\bibliography{papers}

\end{document}